\documentclass[aps,pra,twocolumn,groupedaddress,showpacs,amsmath,amssymb,amsfonts]{revtex4}

\usepackage{graphicx}

\usepackage{color}
\definecolor{gray}{rgb}{0.7,0.7,0.7}

\renewcommand{\v}[1]{\ensuremath{\mathbf{#1}}} % for vectors
\renewcommand{\k}{\v{k}}
\newcommand{\avg}[1]{\left< #1 \right>} % for averagef
\newcommand{\abs}[1]{\left\vert #1 \right\vert} % for absolute value
\newcommand{\nn}{\nonumber}
\newcommand{\ket}[1]{\left| #1 \right>} % for Dirac bras
\renewcommand{\a}{\alpha}

\begin{document}

\title{Stability of three-sublattice order in $S=1$ bilinear-biquadratic
Heisenberg Model on anisotropic triangular lattices}

\author{Yu-Wen Lee}
%\email{ywlee@thu.edu.tw} %
\affiliation{Department of Physics, Tunghai University, Taichung
40704, Taiwan}

\author{Yung-Chung Chen}
\affiliation{Department of Physics, Tunghai University, Taichung
40704, Taiwan}

\author{Min-Fong Yang}
%\email{mfyang@thu.edu.tw} %
\affiliation{Department of Physics, Tunghai University, Taichung
40704, Taiwan}

\date{\today}

\begin{abstract}
The $S=1$ bilinear-biquadratic Heisenberg model on anisotropic
triangular lattices is investigated by several complementary
methods. Our focus is on the stability of the three-sublattice
spin nematic state against spatial anisotropy. We find that,
deviated from the case of isotropic triangular lattice, quantum
fluctuations enhance and the three-sublattice spin nematic order
is reduced. In the limit of weakly coupling chains, by mapping the
systems to an effective one-dimensional model, we show that the
three-sublattice spin nematic order develops at infinitesimal
interchain coupling. Our results provide a complete picture for
smooth crossover from the triangular-lattice case to both the
square-lattice and the one-dimensional limits.
\end{abstract}


\maketitle

%%%%%%%%%%%%%%%%%%%%%%%%%%%%%%%%%%%%%%%%%%%%%%%%%%%%%%%%%%%%%%%%%%
\section{introduction}

Spin nematic states are the states of quantum spin systems in
which no spin-dipolar ordering exists, but spin-rotation symmetry
is spontaneously broken due to the appearance of spin-quadrupolar
order.~\cite{Penc_Laeuchli} Prominent examples for the existence
of such spin nematic phases include the spin-1
bilinear-biquadratic (BLBQ)
model.~\cite{Penc_Laeuchli,Papanicolaou88}
Interest in quantum states with spin nematic order has been raised
recently by experimental findings in NiGa$_2$S$_4$, which is an
insulating quantum magnet with spin-1 Ni$^{2+}$ ions living on a
triangular lattice.~\cite{Nakatsujui05} This system is found to be
in a gapless ground state without spin-dipolar ordering. It has
been suggested that this compound can be considered as a physical
realization of the BLBQ model on a triangular lattice and the
candidate ground state is characterized by a three-sublattice spin
nematic order.~\cite{Tsunetsugu06,Lauchli06} The observed gapless
excitation spectrum thus corresponds to the Nambu-Goldstone modes
associated with spontaneous breaking of spin-rotation symmetry.

\begin{figure}
\includegraphics[clip,width=0.95\columnwidth]{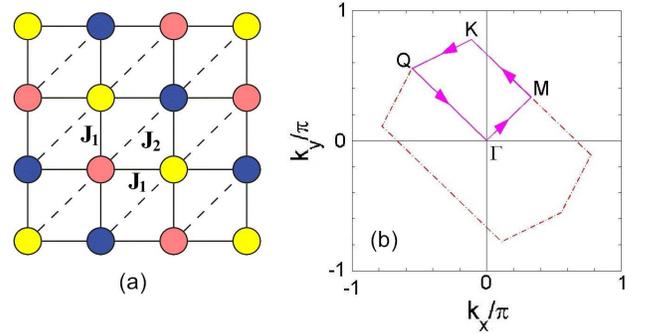}
\caption{(Color online) %
(a) Illustration of the anisotropic triangular lattice in a square
topology and the schematic representation of three-sublattice spin
nematic order. Two groups of interactions, $J_1$ and $J_2$, are
denoted by solid and dashed links, respectively. Here the three
mutually orthogonal vectors $\mathbf{d}_i$ in the mean-field
analysis (see Sec.~\ref{LFW}) are associated with three different
colors.
(b) Brillouin zone of square lattice. The reduced Brillouin zone
for the three-sublattice order is enclosed by dashed lines. The
$\k$ path used in Fig.~\ref{fig:dispersion} is defined as follows:
$\Gamma:(0,0)$, $M:(\pi/3,\pi/3)$, $K:(-\pi/9,7\pi/9)$, and
$Q:(-5\pi/9,5\pi/9)$. }\label{fig:lattice}
\end{figure}

While consensus has been reached for the ground states of the BLBQ
model on a triangular lattice,~\cite{Penc_Laeuchli} physics for
spatially anisotropic models has not yet been addressed. Here we
consider the spin-1 BLBQ model on anisotropic triangular lattices
[see Fig.~\ref{fig:lattice}(a)] defined by the Hamiltonian,
\begin{eqnarray}
  H &=& J_1 \sum_{\langle i,j\rangle}
  \left[ \cos\theta\; \mathbf{S}_{i} \cdot \mathbf{S}_{j}
    + \sin\theta\; (\mathbf{S}_{i} \cdot \mathbf{S}_{j})^{2} \right] \nonumber \\
    & & + J_2 \sum_{\langle\langle i,j \rangle\rangle}
  \left[ \cos\theta\; \mathbf{S}_{i} \cdot \mathbf{S}_{j}
    + \sin\theta\; (\mathbf{S}_{i} \cdot \mathbf{S}_{j})^{2}
    \right] \; , \label{eq:JKModel}
\end{eqnarray}
where $\mathbf{S}_{i}$'s are spin-1 operators. We use the
notations $\langle i,j\rangle$ and $\langle\langle i,j
\rangle\rangle$ to denote the nearest-neighbour bonds and the
bonds along only one of the diagonals, respectively. $J_1$ and
$J_2$ are the coupling strengths on the corresponding bonds. The
relative strength of the linear and the biquadratic couplings is
parameterized by $\theta$. Here $\alpha\equiv J_2/J_1$ defines the
extent of spatial anisotropy. As the anisotropy $\alpha$ increases
from zero, the model changes from the square lattice to the
isotropic triangular lattice, and eventually to decoupled chains.
We concentrate on the parameter region of $J_1,\;J_2\geq0$ and
$\pi/4\leq\theta<\pi/2$, where the ground states with
three-sublattice spin nematic order are expected.

%------------------  figure  --------------------------------
%------------------  figure  --------------------------------

The model in Eq.~\eqref{eq:JKModel} includes several limiting
cases, in which the ground states are known:
\par (i) At $J_1=0$, the anisotropic model becomes a set of decoupled
one-dimensional (1D) BLBQ spin chains, in which each spin
interacts with two neighbors only (i.e., the coordination number
$z=2$). For each spin chain with $\pi/4\leq\theta<\pi/2$, the
system is found to be in an extended critical phase with soft
modes at momenta $k=0$, $\pm2\pi/3$.~\cite{Fath91} Away from the
SU(3) point ($\theta=\pi/4$), this phase develops dominant
antiferro-quadrupolar correlations with a period of three lattice
units (i.e., almost ``trimerized" ground
state).~\cite{Itoi97,Lauchli06_PRB}
\par (ii) At $J_1=J_2$, the model in Eq.~\eqref{eq:JKModel} is
equivalent to an isotropic triangular-lattice model with $z=6$.
The ground state for $\pi/4\leq\theta<\pi/2$ is shown to poss a
three-sublattice spin nematic order,~\cite{Tsunetsugu06,Lauchli06}
where the nematic directors on the three sublattices $A$, $B$, and
$C$ of the triangular lattice are orthogonal to each other (say,
along $\hat{x}$, $\hat{y}$, and $\hat{z}$, respectively). The
schematic representation of this order is shown in
Fig.~\ref{fig:lattice}(a).
\par (iii) At $J_2=0$, our model reduces to a square-lattice model with
$z=4$. It is established only recently that the ground state for
$\pi/4\leq\theta<\pi/2$ develops an unexpected three-sublattice
spin nematic order as a consequence of a subtle quantum
order-by-disorder mechanism.~\cite{Toth10,Toth12,Bauer12}

In this paper, the spatially anisotropic BLBQ model in
Eq.~\eqref{eq:JKModel} is investigated. We pay our attention to
the effect of spatial anisotropy on the stability of the
three-sublattice spin nematic state in this model.
Our main results for generic cases of anisotropy are based on the
linear flavor-wave (LFW)
theory,~\cite{Penc_Laeuchli,Papanicolaou88,Chubukov90,Joshi99}
which has been applied to the triangular-lattice as well as the
square-lattice cases with
success.~\cite{Tsunetsugu06,Lauchli06,Toth10,Toth12,Bauer12} We
find that three-sublattice spin nematic order is most robust in
the case of isotropic triangular lattice with anisotropy
$\alpha\equiv J_2/J_1 = 1$. As deviated from this $\alpha=1$ case,
quantum fluctuations enhance and the order is reduced. This
behavior is reasonable, since the coordination number $z$ is
decreased both in the $\alpha\rightarrow 0$ (square-lattice limit)
and the $\alpha\rightarrow \infty$ (decoupled-chain limit) cases,
and stronger quantum fluctuations are thus allowed.
In order to address the validity of the LFW theory, we have
performed exact diagonalizations (ED) on lattices of small sizes.
By comparing our LFW predictions specialized to finite-size
systems with the numerical results, we find that quantum
fluctuations obtained by the LFW theory are overestimated,
especially in both $\alpha\rightarrow 0$ and $\alpha\rightarrow
\infty$ limits. Thus the stability region of the three-sublattice
state could be larger than the LFW predictions.
Since previous numerical
investigations,~\cite{Toth10,Toth12,Bauer12} have shown nonzero
order in the square-lattice case at $\theta=\pi/4$, one may expect
that the three-sublattice order could persist down to the
$\alpha=0$ limit in the whole region of $\pi/4\leq\theta<\pi/2$.
In the opposite large-$\alpha$ limit, the status is much less
clear. Because there is no true long-range order at the
decoupled-chain limit
($J_1/J_2=1/\alpha=0$),~\cite{Fath91,Itoi97,Lauchli06_PRB} an
interesting issue is whether a nonzero interchain coupling is
necessary or not for the appearance of two-dimensional (2D)
three-sublattice order. By mapping from the system of weakly
coupled chains ($J_1/J_2\ll 1$) to an effective 1D model, we show
that the critical value of interchain coupling is $(J_1/J_2)_c=0$
for all $\pi/4<\theta<\pi/2$. In other words, the transition from
the 2D three-sublattice phase to the 1D ``trimerized" critical
phase~\cite{Fath91,Itoi97,Lauchli06_PRB} should occur at infinite
$\alpha$.

The rest part of the paper is organized as follows. Generic cases
of anisotropy are discussed in Sec.~II, where details of the LFW
analysis are presented in Sec.~II~A and the comparison between the
finite-size LFW and the ED results is made in Sec.~II~B. The case
in the decoupled-chain limit is explored in Sec.~III through
field-theoretical approach as well as ED calculations. The last
section is devoted to our conclusions.

%%%%%%%%%%%%%%%%%%%%%%%%%%%%%%%%%%%%%%%%%%%%%%%%%%%%%%%%%%%%%%%%%%
\section{Generic anisotropy}
\subsection{linear flavor-wave analysis} \label{LFW}

The LFW theory starts from representing the model in
Eq.~\eqref{eq:JKModel} in terms of three-flavor Schwinger bosons
$a_{i,\alpha}$ under the local constraint $\sum_{\alpha}
a_{i,\alpha}^\dagger
a_{i,\alpha}=1$.~\cite{Penc_Laeuchli,Papanicolaou88,Chubukov90,Joshi99,Tsunetsugu06,Lauchli06,Toth10,Toth12,Bauer12}
The Schwinger bosons $a_{i,\alpha}^\dagger$ (with $\alpha=x$, $y$,
$z$) create
three time-reversal-invariant local basis states, %
$|x\rangle = \frac{i}{\sqrt{2}}(|s_z=1\rangle - |s_z=-1\rangle)$, %
$|y\rangle = \frac{1}{\sqrt{2}}(|s_z=1\rangle + |s_z=-1\rangle)$, and %
$|z\rangle = -i\;|s_z=0\rangle$. %
In terms of these bosons, the spin operators become %
$S_i^\alpha=-i\sum_{\beta,\gamma}\epsilon_{\alpha\beta\gamma}a_{i,\beta}^\dagger
a_{i,\gamma}$.
We denote $\v{d}_i$ as the local ordering vector and let
$\{\v{d}_i, \v{e}_i, \v{f}_i\}$ forming a local orthonormal basis.
A generic local quantum state can be represented by the linear
combination of the three basis states $\{\ket{\v{d}_i},
\ket{\v{e}_i}, \ket{\v{f}_i}\}$. Let $a_i^\dagger$, $b_i^\dagger$
and $c_i^\dagger$ representing the Schwinger boson operators which
create the local states $\{\ket{\v{d}_i}, \ket{\v{e}_i},
\ket{\v{f}_i}\}$ out of the Schwinger boson vacuum. They are
related to the operators $a_{i,\a}$ through the relation
$a_{i,\a}=d_{i,\a}\,a_i+e_{i,\a}\,b_i+f_{i,\a}\,c_i$. Within the
LFW analysis, we solve the local constraint $a_i^\dagger a_i
+b_i^\dagger b_i +c_i^\dagger c_i=1$ by replacing
$a_i=\sqrt{1-b_i^\dagger b_i-c_i^\dagger c_i}\approx 1$ and thus
$a_{i,\a}\approx d_{i,\a}+e_{i,\a}\,b_i+f_{i,\a}\,c_i$.
It has be shown that, for $\pi/4\leq\theta<\pi/2$, the mean-field
energy of the nearest-neighbor bond is minimized when the
$\v{d}_i$ vectors are mutually orthogonal.~\cite{Penc_Laeuchli} In
the case of isotropic triangular lattice, these $\v{d}_i$ vectors
are given by the unit vectors along the $x$, $y$, and $z$
directions on the three sublattices [see
Fig.~\ref{fig:lattice}(a)]. Employing this mean-field condition
and the approximate expression for $a_{\a,i}$, the model in
Eq.~\eqref{eq:JKModel} reduces to the following quadratic LFW
Hamiltonian (up to a constant term),
\begin{align}
H_\textrm{LFW}=2\sum_\k &\left[\epsilon_0 (b_\k^\dagger b_\k
+c_\k^\dagger c_\k)
+(\Delta_\k^* b_\k c_{-\k} + \textrm{h.c.}) \right.\nn\\
&\left.+(\phi_\k b_\k^\dagger c_\k + \textrm{h.c.} )\right] \, .
\label{eq:FW_Hamiltonian}
\end{align}
Here the values of $\k$ run over the first Brillouin zone of the
square lattice and
\begin{eqnarray}\label{eq:Delta_k}
\epsilon_0&=&(J_1+\frac{J_2}{2})\sin\theta \; , \nn\\
\Delta_\k&=&\frac{\cos\theta}{2}\left[J_1(e^{ik_x}+e^{ik_y})+J_2\;e^{-i(k_x+k_y)}\right] \; , \\
\phi_\k&=&(\tan\theta-1)\Delta_\k \; . \nn
\end{eqnarray}
For $\mathbf{k}\neq0,\pm \mathbf{k}_0$ with $\mathbf{k}_0=(2\pi/3,
2\pi/3)$, the resulting quadratic bosonic Hamiltonian can be
diagonalized by the Boguliubov transformation, and the
corresponding excitation spectrums are given by
\begin{align}
\omega_{1,2}(\k)&= 2\left[\epsilon_0^2+\abs{\phi_\k}^2-\abs{\Delta_\k}^2 \right.\nn\\
& \left.
\pm\sqrt{2\abs{\phi_\k}^2(2\epsilon_0^2-\abs{\Delta_\k}^2)
+(\phi_\k^*)^2\Delta_\k^2+(\Delta_\k^*)^2\phi_\k^2}\right]^{1/2}\,
. \label{eq:FW_excitations_spectrums}
\end{align}
The modes for $\mathbf{k}=0,\pm \mathbf{k}_0$ cannot be
diagonalized in this way, because the Boguliubov transformation
becomes singular here. As discussed in the finite-size spin-wave
theory for spin-$1/2$ Heisenberg model,~\cite{Zhong93,Trumper00}
these singular modes have no contribution to the ground-state
energy, while removal of these modes is required in the
computation of order parameter.

Some general features of the LFW excitation spectrums are
described below. At the SU(3) point of $\theta=\pi/4$, we have
$\phi_\k=0$ and thus these two excitation modes become degenerate
in energy. Away from this special point, $\omega_1(\mathbf{k})$
gives a gapped mode, while $\omega_2(\mathbf{k})$ is gapless and
has nodes at $\mathbf{k}=0,\pm \mathbf{k}_0$. We remind that the
primitive unit cell of three-sublattice states contains three
lattice sites as its basis and therefore its size becomes three
times larger. As a consequence, the original Brillouin zone of
square lattice behaves as an extended Brillouin zone [as shown in
Fig.~\ref{fig:lattice}(b)], such that each branch of excitations
in Eq.~\eqref{eq:FW_excitations_spectrums} becomes three-fold
degenerate within the original Brillouin zone.
As a simple check for our derivations, we point out that the
obtained dispersions at $J_2=0$ do reduce to those in
Ref.~\onlinecite{Toth10} for the square-lattice case. For the case
of isotropic triangular lattice ($J_1=J_2$), they are equivalent
to the results in Ref.~\onlinecite{Tsunetsugu06}.

To determine the stability region of the three-sublattice states,
a suitable order parameter should be measured. In the spin nematic
state, spin rotational symmetry is spontaneously broken, though
time reversal symmetry is preserved. In such a state, average
magnetic moment must vanish ($\langle \mathbf{S} \rangle =0$).
Nevertheless, quadrupole order can appear, which is characterized
by a nonzero expectation value of the symmetric and traceless
rank-2 tensor operator
\begin{equation}
{\cal Q}^{\alpha \beta}_i = \frac{1}{2} \left( S^\alpha_i
S^\beta_i + S^\beta_i S^\alpha_i \right) - \frac{2}{3}
\delta^{\alpha \beta} \; . \label{eq:nematic}
\end{equation}
Here $S^\alpha_i$ is the $\alpha$ component of spin-1 operator at
site $i$ and $\delta^{\alpha \beta}$ is the Kroneker delta symbol.
In terms of the local ordering vector $\mathbf{d}_i$, the
expectation value of this quadrupole operator can be written as
\begin{equation}
\langle{\cal Q}_{i}^{\alpha\beta}\rangle = - q \left( d_i^\alpha
d_i^\beta -\frac{1}{3}\delta^{\alpha\beta} \right) \; ,
\label{eq:nematic_exp}
\end{equation}
where the constant value of $q$ describes the magnitude of the
quadrupolar ordering. For both cases of the triangular and the
square lattices,~\cite{Tsunetsugu06,Toth10} the $\mathbf{d}_i$
vectors of the three-sublattice states point along three
orthogonal directions in three different sublattices, as shown in
Fig.~\ref{fig:lattice}(a). From Eqs.~\eqref{eq:nematic} and
\eqref{eq:nematic_exp}, we have (from now on, summation is implied
over the repeated \emph{Greek} indices)
\begin{equation}
q = - \frac{3}{2} \langle{\cal Q}_{i}^{\alpha\beta}\rangle
d_i^\alpha d_i^\beta = 1 - \frac{3}{2} \langle
(\mathbf{S}_i\cdot\mathbf{d}_i)^2 \rangle \; .
\label{eq:order_parameter}
\end{equation}
At classical level, $q=1$ because
$(\mathbf{S}_i\cdot\mathbf{d}_i)|\mathbf{d}_i\rangle=0$.

%------------------  figure  --------------------------------
\begin{figure}
\includegraphics[clip,width=0.9\columnwidth]{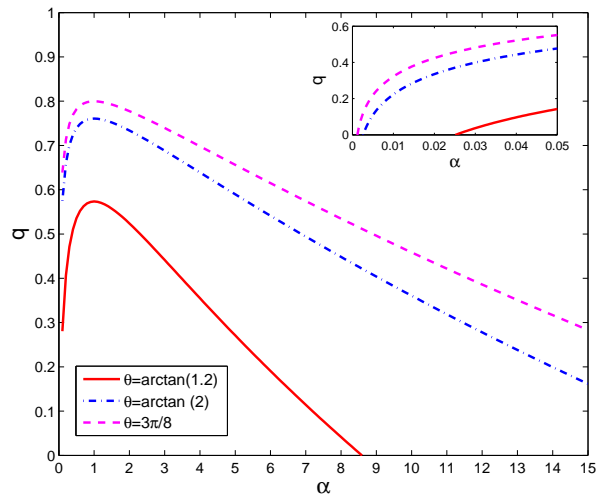}
\caption{(Color online) %
Effect of anisotropy $\alpha=J_2/J_1$ on the quadrupole order
parameter $q$ within the LFW theory for various values of
$\theta$.
The inset shows the details around the region of $\alpha=0$. }
\label{fig:order}
\end{figure}
%------------------  figure  --------------------------------

\begin{figure}
\includegraphics[clip,width=0.9\columnwidth]{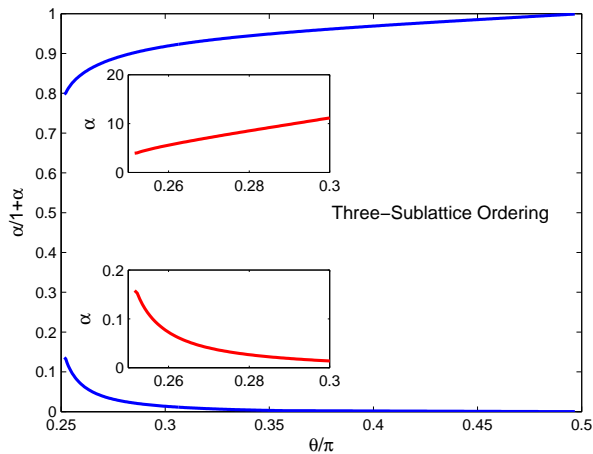}
\caption{(Color online) %
Phase diagram for the spatially anisotropic $S=1$ BLBQ model in
Eq.~\eqref{eq:JKModel} determined by the LFW theory.
The insets show the details of two phase boundaries around
$\theta=\pi/4$. } \label{fig:phase-diag}
\end{figure}

Within the LFW theory, $q$ can be expressed by
\begin{equation}
q=1-\frac{3}{2}\avg{\Delta n_a} \label{eq:FW_order_parameter}
\end{equation}
with
\begin{equation}
\avg{\Delta n_a}=1-\frac{1}{N}\sum_i \avg{a_i^\dagger a_i} =
\frac{1}{N}\sum_i \avg{b_i^\dagger b_i+c_i^\dagger c_i}
\end{equation}
being the deviation of the number density for the Schwinger boson
$a_i$ from its classical value of one. Here $N$ is the total
number of lattice sites. We note that this expression of $q$ is
nothing but the local moment defined in Eq.~(24) of
Ref.~\onlinecite{Bauer12}. From this expression, it is obvious
that the quantum correction for $q$ comes from the non-vanishing
contribution of $\avg{\Delta n_a}$. When $\avg{\Delta n_a}$
increases to $2/3$, $q$ vanishes. This gives a phase transition
out of the three-sublattice states within the LFW theory.
The explicit expression of $\avg{\Delta n_a}$ is given by
\begin{equation}
\avg{\Delta n_a} = \frac{1}{N}\sum_{\k\neq0,\pm \mathbf{k}_0}
\left(\frac{\epsilon_0+\abs{\phi_\k}}{\omega_1(\k)}
+\frac{\epsilon_0-\abs{\phi_\k}}{\omega_2(\k)}-1\right) \; .
\label{eq:Delta_na}
\end{equation}
Here the singular modes at $\mathbf{k}=0,\pm \mathbf{k}_0$ are
excluded from this summation.
Note that, for the case of square lattice ($J_2=0$) and at the
SU(3) point ($\theta=\pi/4$), Eq.~\eqref{eq:Delta_na} reduces to
$\avg{\Delta n_a} = \frac{1}{N} \sum_{\k\neq0,\pm \mathbf{k}_0}
\left(\frac{1}{\sqrt{1-\abs{\gamma_\k}^2}}-1\right)$ with
$\gamma_\k=\cos[(k_x-k_y)/2]$, and reproduces the previous result
(see Eq.~(22) of Ref.~\onlinecite{Bauer12}). The general behavior
of the order parameter $q$ as functions of $\alpha$ for distinct
values of $\theta$ is shown in Fig.~\ref{fig:order}. It is seen
clearly that the three-sublattice nematic order is most robust at
$\alpha=1$. Far away from this point of isotropic triangular
lattice, quantum fluctuations arising from the flavor-wave
excitations become more stronger, and they destroy eventually the
quadrupolar ordering in both limits of $\alpha\rightarrow 0$ and
$\alpha\rightarrow\infty$.
Exploiting Eqs.~\eqref{eq:FW_order_parameter}
and~\eqref{eq:Delta_na} and employing the condition $q=0$ as the
criterion for the transitions out of the three-sublattice states,
we can establish the phase boundaries of the three-sublattice
states as shown in Fig.~\ref{fig:phase-diag}. We find that, in
general, the lower transition points are nonzero and the upper
ones are large but finite. The stability region of the
three-sublattice order is largely reduced as $\theta$ approaches
the SU(3) point ($\theta=\pi/4$). It implies that there exist more
low-lying excitations and thus larger quantum fluctuations as
$\theta$ gets closer to $\pi/4$.

%------------------  figure  --------------------------------
%------------------  figure  --------------------------------

\begin{figure}
\includegraphics[clip,width=0.9\columnwidth]{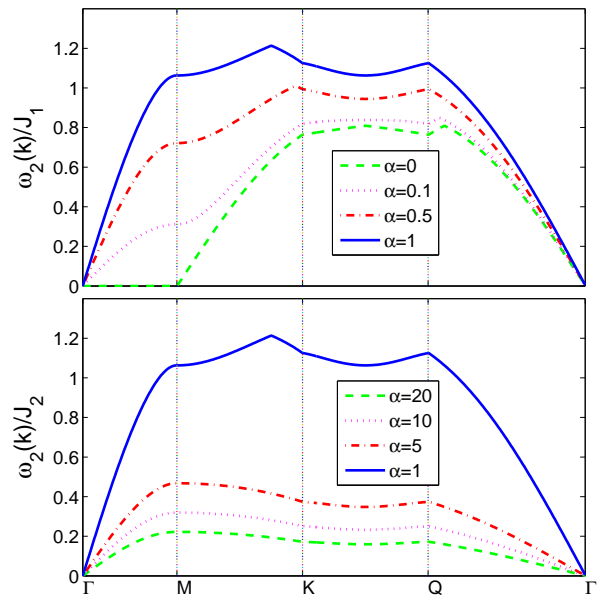}
\caption{(Color online) %
Dispersion relation of gapless branch $\omega_2(\mathbf{k})$ of
the flavor-wave excitation for different anisotropy parameters
$\alpha=J_2/J_1$ at $\theta=0.3\pi$. Note that energies are
measured in unit of $J_1$ for $\alpha<1$ (upper panel), while they
are measured in unit of $J_2$ for $\alpha >1$ (lower panel). }
\label{fig:dispersion}
\end{figure}

As seen from the expression of Eq.~\eqref{eq:Delta_na}, the
gapless mode $\omega_2(\mathbf{k})$ should make a dominant
contribution in reducing the three-sublattice order. To have a
better understanding of the enhancement of quantum fluctuations
both in the square-lattice limit ($\a\to 0$) and the quasi-1D
limit ($\a\gg 1$), it is instructive to examine the softening
behavior of this excitation mode more closely.
The flavor-wave dispersions of the gapless branch
$\omega_2(\mathbf{k})$ along the path defined in
Fig.~\ref{fig:lattice}(b) for various values of anisotropy
$\alpha=J_2/J_1$ are shown in Fig.~\ref{fig:dispersion}. It can be
seen that, as system approaches the square-lattice limit
($\alpha\to 0$), the flavor-wave velocity at the $\Gamma$ point
(defined by the slope of the dispersion relation) decreases to
zero. Therefore, the excitation modes along the $\Gamma$-$M$ line
(i.e., line of $k_x=k_y$) become zero-energy modes eventually. On
the other hand, in the limit of decoupled chains
($\alpha\to\infty$), the excitation energies in the whole
Brillouin zone are softened. Within the LFW theory, the
disappearance of three-sublattice order in both limits of
$\alpha\to 0$ and $\alpha\to\infty$ can be explained by such
softening in energy.

%------------------  figure  --------------------------------
%------------------  figure  --------------------------------

Near the node, say, at $\k=0$, analytic expressions can be
derived. By expanding $\Delta_\k$ in Eq.~\eqref{eq:Delta_k} near
$\k=0$, we can show that the gapless flavor-wave mode behaves as
$\omega_2(\mathbf{k})/J_1 \approx \sqrt{c_+^2 k_+^2+c_-^2 k_-^2}$
with $k_\pm=\frac{1}{\sqrt{2}}(k_x\pm k_y)$,
$c_+=\frac{3}{\sqrt{2}}\sqrt{\sin(2\theta)\a}$, and
$c_-=\frac{1}{\sqrt{2}}\sqrt{\sin(2\theta)(\a+2)}$. For $\a\to 0$,
we have $c_{+}\to 0$ while $c_{-}$ remaining finite. The outcome
of $c_{+}=0$ for $\alpha=0$ gives a nodal line in the excitation
spectrum along the $k_x=k_y$ direction (i.e., $\Gamma$-$M$ line),
as observed in Fig.~\ref{fig:dispersion}. Within the LFW theory,
these soft modes play a significant role in the destruction of the
nematic order in the square-lattice limit, as noticed in the
previous investigations.~\cite{Toth10,Bauer12}
On the other hand, in the extreme anisotropic quasi-1D limit
($J_1\to 0$ or $\a\to\infty$), we have $c_-/c_+=1/3$ for all
values of $\theta$. That is, this ratio of the flavor-wave
velocities does not goes to zero in the quasi-1D limit. Instead,
the whole spectrum, in unit of the diagonal coupling $J_2$,
becomes nearly flat in the entire Brillouin zone, as seen from
Fig.~\ref{fig:dispersion}. Therefore, the associated quantum
fluctuations become more and more significant and finally the 2D
nematic order ceases to exist for $\a$ being large enough.
We stress that the mode softening of the flavor-wave excitations
in the quasi-1D limit is quite different from what one got for the
spin-wave excitations in spatially anisotropic spin-$1/2$
antiferromagnetic Heisenberg models on either
triangular~\cite{Merino99} or square
lattices.~\cite{Sakai89,Affleck94} In these spin-$1/2$ cases, one
can show that, as the interchain couplings approach zero, only the
spin-wave velocity $c_\perp$ for the excitation transverse to the
chains will vanish, but the spin-wave velocity $c_{\parallel}$ for
the excitation along the chains will remain finite. Thus the whole
spectrum, in unit of the intrachain coupling, never becomes nearly
flat in the whole Brillouin zone, and the ratio of these two
spin-wave velocities $c_\perp/c_\parallel$ does go to zero in the
quasi-1D limit.

We remind that the LFW analysis is valid only when quantum
fluctuations are weak (i.e., only when $\langle b_i^\dagger
b_i\rangle + \langle c_i^\dagger c_i\rangle\ll 1$) because the
local constraint, $a_i^\dagger a_i +b_i^\dagger b_i +c_i^\dagger
c_i=1$, is considered only approximatively. Therefore, we should
be cautious with the LFW results about the phase boundaries shown
in Fig.~\ref{fig:phase-diag}, since large quantum fluctuations are
expected near the transition points. To examine the validity of
the LFW predictions, comparison with exact results is necessary.

%%%%%%%%%%%%%%%%%%%%%%%%%%%%%%%%%%%%%%%%%%%%%%%%%%%%%%%%%%%%%%%%%%
\subsection{exact diagonalizetion}

In this subsection, we perform ED calculations for the Hamiltonian
in Eq.~\eqref{eq:JKModel} on small clusters and compare the
results with those obtained by the LFW analysis on the same
clusters.

We remind that there exist subtleties in making careful comparison
of order parameter between symmetry-breaking solutions (say, LFW
results) and symmetry-nonbreaking ones (say, ED findings). Such an
observation has been put forward in Ref.~\onlinecite{Bernu94} in
concern with magnetic ordering in spin-$1/2$ Heisenberg
antiferromagnets on a triangular lattice.
To uncover the long-range order on lattices of small sizes, a
proper quantity has to be measured in ED calculations. Here the
squared quadrupole moment $\mathcal{Q}^2$ in a given sublattice
(say, $A$ sublattice) is considered,
\begin{equation}
\mathcal{Q}^2 %
\equiv \left\langle \left(\sum_{j\in A}\mathcal{Q}^{\alpha \beta}_j \right)^2 \right\rangle %
= \sum_{i,j\in A} \left\langle \mathcal{Q}^{\alpha \beta}_i \mathcal{Q}^{\alpha \beta}_j \right\rangle \; . %
\label{eq:sub-quad}
\end{equation}
As mentioned before, the Einstein summation convention for the
repeated \emph{Greek} indices is assumed. The signature of
three-sublattice order will be manifested as a macroscopic value
of $\mathcal{Q}$.
To obtain a order parameter that is normalized to 1 in the absence
of quantum fluctuations, the sublattice quadrupole moment
$\mathcal{Q}$ should be divided by a size-dependent normalization
factor. For deriving the correct normalization factor, it should
be kept in mind that the sublattice quadrupole moment cannot be
treated as a classical quantity. For example, it can be shown that
there exists an exact operator identity: $\mathcal{Q}^{\alpha
\beta}_i \mathcal{Q}^{\alpha \beta}_i=5/3$ on a given site $i$. On
the other hand, when $i$ and $j$ denote different sites of the
same sublattice, $\left\langle \mathcal{Q}^{\alpha \beta}_i
\mathcal{Q}^{\alpha \beta}_j \right\rangle=2/3$ for fully aligned
classical ordered state. From these observations, the maximum
quantum value of $\mathcal{Q}^{2}$ can be shown to be
$(2N^2/27)(1+9/2N)$ for systems of $N$ sites. Thus a valid
definition of the order parameter would be
\begin{equation}
q = \sqrt{\frac{27\mathcal{Q}^{2}}{2N^2(1+\frac{9}{2N})}} \; .
\label{eq-order}
\end{equation}
Now $q$ saturates at one in the classical state and should be
decreased by quantum fluctuations in the quantum ground state.
We stress that, for careful comparison between the ED and the LFW
results for small values of $N$, it is important to use the
correct normalization factor in the definition of the order
parameter $q$.

%------------------  figure  --------------------------------
\begin{figure}
\includegraphics[clip,width=0.9\columnwidth]{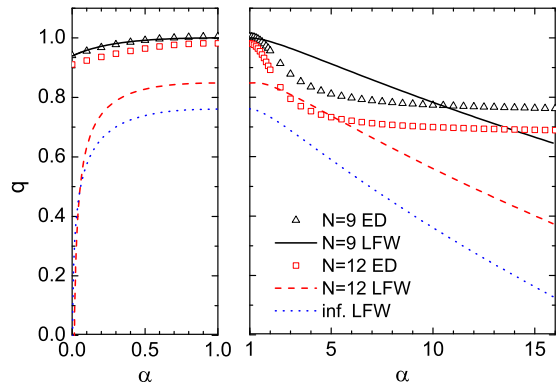}
\caption{(Color online) %
Comparison of order parameter $q$ between the ED and the LFW
results for different lattice sites as anisotropy $\alpha$ ia
varied. Here, $\theta=\arctan(2)$. The ED data are obtained on
lattices with sites $N=9$ (triangles) and $N=12$ (squares). The
LFW results of the corresponding sizes are shown by solid and
dashed lines, respectively. The lowest curve refers to the
infinite-size LFW results (dotted line). } \label{fig:order_ED}
\end{figure}
%------------------  figure  --------------------------------

The comparison between the ED and the LFW results is shown in
Fig.~\ref{fig:order_ED} for $\theta=\arctan(2)$. The ED data are
calculated by using Eqs.~\eqref{eq:sub-quad} and~\eqref{eq-order}.
On the other hand, the LFW results for systems of finite sizes are
obtained from Eqs.~\eqref{eq:FW_order_parameter}
and~\eqref{eq:Delta_na} by summing the momenta (except
$\mathbf{k}=0$, $\pm \mathbf{k}_0$) determined by the clusters in
ED calculations.
We find that, around the isotropic point ($\alpha=1$), two sets of
results do not differ by large amounts. Good agreement can persist
even down to the $\alpha=0$ limit for the special case of $N=9$.
This indicates that the LFW theory does in general provide good
approximation around $\alpha=1$. However, serious reduction in the
LFW results of $q$ as compared to the ED ones is observed both
when $\alpha\ll 1$ and $\alpha\gg 1$. This implies that quantum
fluctuations are significantly overestimated in the LFW analysis
in both of the square-lattice and the quasi-1D limits. In other
words, the softening of the flavor-wave excitations in both limits
(see Fig.~\ref{fig:dispersion}) should be exaggerated. Thus one
has to go beyond the LFW approximation to find improvements on the
excitation spectrums.
According to previous numerical
investigations,~\cite{Toth10,Toth12,Bauer12} where nonzero order
was reported in the square-lattice case at $\theta=\pi/4$, the
lower phase boundary obtained within the LFW theory (see
Fig.~\ref{fig:phase-diag}) may be illusive. Instead, the
three-sublattice order may persist down to the $\alpha=0$ limit in
the whole region of $\pi/4\leq\theta<\pi/2$.

The above comparison suggests as well that the upper phase
boundary may take much larger values than the ones estimated by
the LFW theory. To achieve the true values of the transition
points in the large $\alpha$ limit, it is instructive to analyze
the model from its quasi-1D limit. Because there is no true
long-range order in strictly 1D models, our main concern is to
show whether a nonzero interchain coupling is necessary or not to
establish the 2D order. This is what we shall do in the next
section.

%%%%%%%%%%%%%%%%%%%%%%%%%%%%%%%%%%%%%%%%%%%%%%%%%%%%%%%%%%%%%%%%%%
\section{weakly-coupled-chain limit}

When $J_1 \ll J_2$, the system described by Eq.~\eqref{eq:JKModel}
reduces to weakly coupled 1D chains with intrachain coupling
strength $J_2$ and weak interchain coupling strength $J_1$. Taking
advantage of conventional mean-field treatment for the interchain
coupling,~\cite{Sakai89,Affleck94,Scalapino75,Schulz96} our
quasi-1D systems can be transformed into effective single-chain
problems.

The desired effective single-chain model can be derived in the
following way. Using the definition of the quadrupole operator in
Eq.~\eqref{eq:nematic}, the part of Hamiltonian with the
interchain coupling $J_1$ can be rewritten as
\begin{equation}
H_1 = %
J_1(\cos\theta-\frac{\sin\theta}{2}) \sum_{\langle ij \rangle} \mathbf{S}_i \cdot \mathbf{S}_j %
+ J_1 \sin\theta \sum_{\langle ij \rangle} {\cal Q}^{\alpha \beta}_i {\cal Q}^{\alpha \beta}_j  %
\label{J1Ham}
\end{equation}
by dropping some constant terms. Again, the repeated \emph{Greek}
indices imply the Einstein summation convention. Assuming the
three-sublattice quadrupolar ordering, in which
$\langle\mathbf{S}_{j}\rangle=0$ but $\langle{\cal
Q}_{j}^{\alpha\beta}\rangle$ is nonzero, $H_1$ can be approximated
by an on-site Hamiltonian,
\begin{equation}
\tilde{H}_1= \sum_n {\cal Q}_n^{\alpha\beta} \left[ J_1 \sin\theta
\sum_m \langle{\cal Q}_{n+m}^{\alpha\beta}\rangle \right] \; ,
\label{eq:onsite_term}
\end{equation}
where $m$ runs over all neighbors on the $J_1$ bonds for the
$n$-th site along a given single chain. Substituting suitable
mean-field expression for $\langle{\cal
Q}_{n+m}^{\alpha\beta}\rangle$, an effective 1D Hamiltonian for
the model in Eq.~\eqref{eq:JKModel} can be written as
\begin{equation}
H_\textrm{eff} = %
J_2 \sum_{n} \left[ \cos\theta\; \mathbf{S}_{n} \cdot
\mathbf{S}_{n+1} + \sin\theta\; (\mathbf{S}_{n} \cdot
\mathbf{S}_{n+1})^{2} \right] + \tilde{H}_1 \; .
\label{eq:1Dmfmodel}
\end{equation}
This describes a 1D BLBQ chain in a self-consistent external field
triggering three-sublattice quadrupolar ordering. For convenience,
we set $J_2\equiv 1$ as the energy unit in this section.
We note that the self-consistent field is proportional to
$J_1\sin\theta$. It is thus expected that, for a given anisotropy
(i.e., for a fixed value of $J_1$), the resulting order will be
stronger as $\theta$ gets closer to $\pi/2$. This observation is
consistent with our LFW results, where the stability region of the
three-sublattice state is pushed toward larger $\alpha=J_2/J_1$ as
$\theta\to\pi/2$.

In the following, both analytical and numerical techniques are
exploited to determine the critical interchain coupling $J_{1,c}$
for the emergence of 2D three-sublattice order.

\subsection{scaling analysis}

Employing the mean-field expression of $\langle{\cal
Q}_{j}^{\alpha\beta}\rangle$ in Eq.~\eqref{eq:nematic_exp}, the
on-site part of Eq.~\eqref{eq:onsite_term} for the effective 1D
Hamiltonian becomes
\begin{equation}
\tilde{H}_1 = h \sum_{n} \left[ 1- \frac{3}{2}
(\mathbf{S}_n\cdot\mathbf{d}_n)^2 \right] \; ,
\label{eq:onsite_term_1}
\end{equation}
where the self-consistent field conjugate to the operator for the
order parameter $q$ in Eq.~\eqref{eq:order_parameter} is defined
by $h\equiv (4/3)qJ_1\sin\theta$.

For $J_1\ll 1$, the effective 1D Hamiltonian in
Eq.~\eqref{eq:1Dmfmodel} can be considered as a 1D BLBQ model in a
weak external field $h$. Thus it should be valid to treat the
effect of $h$ as a perturbation.
It is known from Ref.~\onlinecite{Itoi97} that, near
$\theta=\pi/4$, the 1D BLBQ spin chain can be described by an
SU(3)$_1$ Wess-Zumino-Witten conformal field theory perturbed by
some marginally irrelevant current-current interactions.
To see whether the three-sublattice order can be induced by
vanishing self-consistent field $h$ or not, we need only to
calculate the scaling dimension $\Delta_h$ of the corresponding
operator in the on-site term and then determine its relevancy. If
$\Delta_h<2$, the on-site term provides a relevant perturbation
and thus the three-sublattice order will be induced by an
infinitesimal $h$. Otherwise, the on-site term becomes irrelevant
and the three-sublattice order can be established only when $h$
exceeds a nonzero critical field $h_c$. In this latter case, we
need to determine $h_c$ numerically by solving this model
explicitly.

In terms of the field-theoretical variables discussed in
Ref.~\onlinecite{Itoi97}, the operator in the on-site term of
Eq.~\eqref{eq:onsite_term_1} takes the form of the primary fields
of an SU($\nu$) Wess-Zumino-Witten model for $\nu=3$. That is, the
scaling dimension $\Delta_h$ of our operator is nothing but that
of those primary fields, which is equal to $1-1/\nu=2/3$ according
to the analysis in Ref.~\onlinecite{Itoi97}. Because of
$\Delta_h<2$, the perturbation caused by the self-consistent field
is strongly relevant.
Since the above scaling argument is essentially independent of the
value of $\theta$, we claim that $h_c=0$ and thus $J_{1,c}=0$ for
$\pi/4< \theta <\pi/2$, even though the field theory in
Ref.~\onlinecite{Itoi97} is derived for $\theta$ close to $\pi/4$.
This implies that, for original quasi-1D anisotropic BLBQ model,
true phase transitions out of the three-sublattice states actually
occur at infinite $\alpha$, rather than at large but finite
$\alpha_c$ as obtained in the LFW analysis.

We can go one step further to establish a nonperturbative relation
between the order parameter $q$ and the weak interchain couplings
$J_1$ by making use of the field-theoretic approach. According to
the standard scaling argument,~\cite{Cardy} the order parameter
$q$ induced by the perturbation of self-consistent field $h$
scales as $q\propto h^{\Delta_h/(2-\Delta_h)}$. Combined with the
self-consistency relation, $h=(4/3)qJ_1\sin\theta$, we get
$q\propto(J_1\sin\theta)^{\Delta_h/[2(1-\Delta_h)]}$. Since
$\Delta_h=2/3$, we conclude that $q\propto J_1\sin\theta$.
Interestingly, this result coincides with the one that is expected
naively from the perturbation theory for original $H_1$ without
taking mean-field approximation. This seems to indicate that it is
possible to study this model in its quasi-1D limit directly from
perturbation theory for $H_1$. Instead of pursuing along this
direction, we shall determine the phase boundary by the numerical
ED method below.

%%%%%%%%%%%%%%%%%%%%%%%%%%%%%%%%%%%%%%%%%%%%%%%%%%%%%%%%%%%%%%%%%%
\subsection{exact diagonalization}

In this subsection, the critical values of the interchain coupling
$J_1$ are estimated by the ED method. Here we follow the treatment
in Ref.~\onlinecite{Sakai89} for quasi-1D Heisenberg
antiferromagnets. Our ED results provide numerical evidences in
supporting the above conclusions based on scaling arguments.

For the sake of ED calculations, we assume here that only the $zz$
component of the expectation value $\langle{\cal
Q}_{j}^{\alpha\beta}\rangle$ is nonzero. Thus the effective 1D
Hamiltonian in Eq.~\eqref{eq:1Dmfmodel} has still spin-rotation
symmetry in the $z$ direction and the total $z$-component spin
remains a conserved quantity. This reduces much computational
effort and thus calculations for systems of large sizes become
available. In consistent with Eq.~\eqref{eq:nematic_exp}, the
explicit form of $\langle{\cal Q}_{j}^{\alpha\beta}\rangle$ is
taken to be $\langle{\cal Q}_{j}^{zz}\rangle =
\langle(S^z_j)^2\rangle -
\frac{2}{3}=-\frac{2}{3}q\cos(\mathbf{Q}\cdot\mathbf{r}_j)$.
Substituting the present mean-field solution to
Eq.~\eqref{eq:onsite_term}, the on-site part of the effective 1D
Hamiltonian becomes
\begin{equation}
\tilde{H}_1 = h \sum_{n} \cos(\frac{4\pi}{3}n) (\mathbf{S}^z_n)^2
\; . \label{eq:onsite_term_2}
\end{equation}
Here the self-consistent field is again given by $h\equiv
(4/3)qJ_1\sin\theta$.

\begin{figure}
\includegraphics[clip,width=0.9\columnwidth]{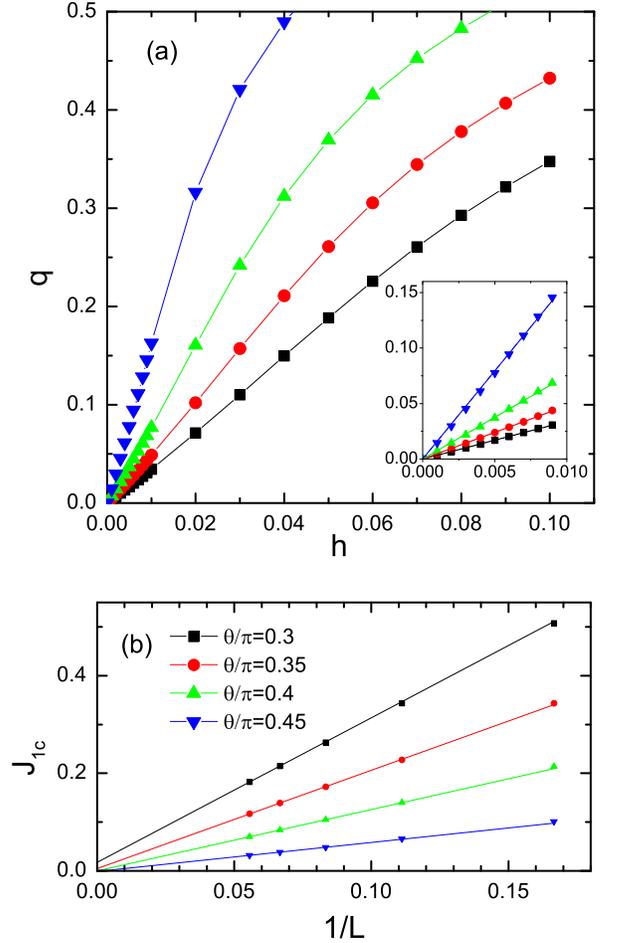}
\caption{(Color online) %
(a) Order parameter $q$ as function of self-consistent field $h$
for various $\theta$'s with $L=18$. Lines are guide to eyes.
Inset: linear fit around $h=0$ region.
(b) Critical value $J_{1,c}$ of the interchain coupling as
function of $1/L$. Lines show the extrapolations in the
thermodynamic limit by using data for the largest two sizes (i.e.,
$L=15$ and 18). } \label{fig:1D_ED}
\end{figure}

By taking $h$ as a free parameter, we diagonalize numerically the
effective single-chain model up to system length $L=18$. For the
present single-chain problem, the order parameter for spin
quadrupole ordering becomes
\begin{equation}
q = -\frac{3}{L}\sum_{n} e^{i \frac{4\pi}{3}n} \langle{\cal
Q}_{n}^{zz}\rangle  \; .
\end{equation}
This expression is compatible with the form of 2D order parameter
used in this subsection. The results of $q$ as function of $h$ for
several $\theta$'s with $L=18$ are presented in
Fig.~\ref{fig:1D_ED}(a). The susceptibility $\chi\equiv(\partial
q/\partial h)|_{h=0}$ can then be evaluated from the slope of the
linear fit as shown in the inset of this figure. Within the
present mean-field approach, to have a nonzero solution of $q$,
the slope $\chi$ of the tangent line around $h=0$ must be larger
than that of the straight line, $q=h/[(4/3)J_1\sin\theta]$, given
by the self-consistent relation. That is, long-range order appears
only when $\chi\geq 1/[(4/3)J_1\sin\theta]$. This requirement
leads to a critical value $J_{1,c}$ of the interchain coupling for
a given length $L$,
\begin{equation}
J_{1,c} = \frac{1}{(4/3)\chi\sin\theta} \; .
\end{equation}
The size dependence of $J_{1,c}$ for various $\theta$'s is shown
in Fig.~\ref{fig:1D_ED}(b). As seen from this figure, size
dependence of $J_{1,c}$ is more prominent as $\theta$ gets closer
to $\pi/4$. This reflects the fact that quantum fluctuations for
the 1D systems become larger as $\theta$ approaches to the SU(3)
point, where more low-energy excitations appear. Except for the
case of $\theta=0.3\pi$, in which size effect may be profound, a
smooth extrapolation of $J_{1,c}$ to zero in the thermodynamic
limit is found for all $\theta$'s. This indicates that the 2D
three-sublattice order will emerge for infinitesimal $J_{1}$
within the whole region of $\pi/4< \theta <\pi/2$. In other words,
the phase transitions out of the three-sublattice states actually
occur at infinite $\alpha$ for original 2D anisotropic BLBQ model.
Thus our ED results lend strong support on the conclusions based
on the scaling arguments discussed in the previous subsection.

%------------------  figure  --------------------------------
%------------------  figure  --------------------------------

\section{Conclusions}

To summarize, we elaborate the effect of spatial anisotropy
$\alpha$ on the stability of the three-sublattice spin nematic
state in the model of Eq.~\eqref{eq:JKModel} through various
analytic as well as numerical approaches. We conclude that the
three-sublattice state is stable for all $0\leq\alpha<\infty$
within the whole region of $\pi/4<\theta<\pi/2$.
Our analysis thus gives a complete picture for smooth crossover
from the triangular-lattice case to both the square-lattice and
the 1D-chain limits as the anisotropy $\alpha$ is varied.
Moreover, our work provides some insights on the validity of the
LFW theory. Basically, the strength of the stability for the
considered order can be understood within the simple LFW theory.
Nevertheless, the predicted phase boundaries and the excitation
spectrums is merely suggestive, especially in both of the
$\alpha=0$ and $\alpha\to\infty$ limits. Because the local
constraints are released in the LFW analysis, it is interesting to
see if great improvement can be obtained through other approaches
(say, the variational Monte Carlo method), in which these
constraints are taken into account rigorously. Such discussions go
beyond the scope of the present work and deserve further
investigations.
%

%%%%%%%%%%%%%%%%%%%%%%%%%%%%%%%%%%%%%%%%%%%%%%%%%%%%%%%%%%%%%%%%%%
\begin{acknowledgments}
Y.-W.L., Y.-C.C., and M.-F.Y. thank the National Science Council
of Taiwan for support under Grant NSC 99-2112-M-029-004-MY3, No.
NSC 99-2112-M-029-002-MY3 and NSC 99-2112-M-029-003-MY3,
respectively.
\end{acknowledgments}
%%%%%%%%%%%%%%%%%%%%%%%%%%%%%%%%%%%%%%%%%%%%%%%%%%%%%%%%%%%%%%%%%%

%---------------------------------------------------------------

\end{document}